\documentclass{article}
\usepackage{ijcai26}

\usepackage{times}
\usepackage{soul}
\usepackage{url}
\usepackage[hidelinks]{hyperref}
\usepackage[utf8]{inputenc}
\usepackage[small]{caption}
\usepackage{graphicx}
\usepackage{amsmath}
\usepackage{amsthm}
\usepackage{booktabs}
\usepackage{algorithm}
\usepackage{algorithmic}
\usepackage[switch]{lineno}

\usepackage{amsfonts,amssymb}

\usepackage{graphicx}
\usepackage{booktabs}
\usepackage{array}
\usepackage{multirow}
\usepackage{makecell}
\usepackage{subcaption}

\usepackage{url}
\usepackage{xcolor}
\usepackage{comment}

\newcommand{\mycomment}[1]{} 

\title{DPDSyn: Improving Differentially Private Dataset Synthesis for Model Training by Downstream Task Guidance }

\author{
	Mingxuan Jia$^1$\and
	Wen Huang$^1$\and
    Weixin Zhao$^1$\and
	Xingyi Wang$^1$\and
    Jian Peng$^1$\And
    Zhishuo Zhang$^2$
    \\
	\affiliations
	$^1$Sichuan University\\
	$^2$Southwest Minzu University\\
	\emails
	\{jiamx, zhaoweixin, wangxingyi97\}@stu.scu.edu.cn,
    \{wen, jianpeng\}@scu.edu.cn,
	476321012@qq.com
}

\begin{document}
	
	\maketitle
	
	\begin{abstract}
	\par How to synthesize a dataset while achieving differential privacy for AI model training is a meaningful but challenging problem. To address this problem, state-of-the-art methods first select original private dataset's multiple low-dimensional distributions that have the potential to approximate the distribution of original private dataset with high precision, and then synthesize a dataset obeying all selected low-dimensional distributions as the synthetic dataset. However, it is difficult to select suitable low-dimensional distributions, which in turn degrades the data utility of resulting synthetic dataset. To improve differentially private dataset synthesis, we propose to train a differentially private AI model for downstream tasks on the original private dataset and utilize the trained model to synthesize datasets. In particular, on the one hand, the AI model satisfies differential privacy so no matter how to use the model does not disclose private information of original private dataset. On the other hand, the AI model is trained to complete the downstream task so the AI model preserves critical information for completing downstream tasks. We utilize the AI model to synthesize datasets to achieve the goal of improving data utility while preserving privacy. Empirical evaluations on four benchmark datasets demonstrate that our proposed DPDSyn consistently outperforms eight state-of-the-art baselines with a maximum improvement of 2.40× in accuracy and 333.73× in synthesis efficiency. Further experiments also validate that DPDSyn has strong scalability across varying data scales. 
        
	\end{abstract}
	
	\section{Introduction}

    \par Dataset publishing contributes to the rapid advancement of AI model training significantly ~\cite{zha2025data}. For example, the publication of ImageNet dataset contributes a lot to the advancement of deep learning \cite{hw_add1}. The publication of 14-million-image ImageNet directly facilitates the dramatic reduction of top-5 classification error rates in the ILSVRC challenge from approximately 26\% to under 3\% \cite{hw_add2} \cite{hw_add3}. The dramatic reduction validates the effectiveness of deep convolutional neural networks. In recent years, the publication of massive datasets drives recent breakthroughs in large language models, such as GPT-5 family and DeepSeek-V3 series \cite{weber2024redpajama} \cite{liu2024sphinx} \cite{wang2025parameters}.
 
    \par The publication of private datasets is the key to further improve AI model training \cite{ye2024openfedllm}. As the contribution of dataset publishing is recognized, many useful non-private datasets are published for public usage, resulting in that the contribution of publishing non-private datasets to the development of AI model training is gradually diminishing. Due to strict privacy regulations, there are almost no publicly accessible private datasets, leading to that private datasets are not utilized to train AI models widely. Thus, the publication of private datasets has the potential to further enhance AI model training.  

    \begin{figure}[!h]
		\centering
        \includegraphics[scale=0.7]{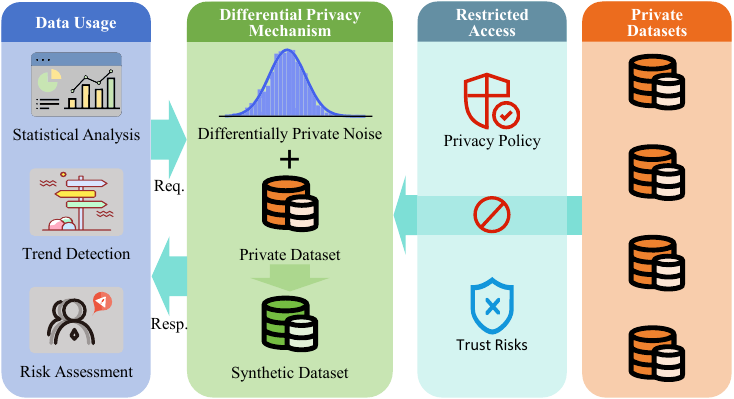}
		\caption{Application Scenario of DP Dataset Synthesis.}
		\label{fig:introduction}
	\end{figure}

    \par Publishing private datasets while preserving privacy could be achieved by DP (differentially private) dataset synthesis as shown in Figure~\ref{fig:introduction}. For example, to publish private datasets while preserving privacy with differential privacy, Alzantot et al. utilized DP-SGD (Stochastic Gradient Descent) to train Wasserstein GAN (Generative Adversarial Networks). The trained GAN is used to generate publicly accessible datasets based on a private dataset needed to be protected \cite{srivastava2019GAN}. Jia et al. proposed ABSyn to publish private datasets with protection of differential privacy~\cite{jia2024absyn}. In particular, they utilized multiple low-dimensional distributions to approximate high-dimensional distributions such that the correlation relationship among data attributions could be preserved as much as possible.

    \par However, existing methods of DP dataset synthesis still have room for further improvement in terms of data utility. Specifically, existing methods of DP dataset synthesis are mainly divided into two categories: GAN-based methods such as \cite{srivastava2019GAN} and select-measure-generate methods such as ABSyn \cite{jia2024absyn}. The GAN-based method tends to suffer from training instability, leading to suboptimal data utility in downstream tasks \cite{garg2024exploring} \cite{ghosheh2024survey}. The select-measure-generate method focus on preserving the correlation relationship among data attributions in the private datasets and the preservation of correlation relationship makes the data utility high in general cases. But, select-measure-generate methods do not take the special characteristics of downstream tasks into account, which leaves us the chance to further improve the data utility of DP dataset synthesis. 
    
    \par In this paper, we utilize the special characteristics of downstream tasks to further improve data utility of DP dataset synthesis. In particular, to capture the key features which can significantly improve performance of downstream tasks, we train an AI model for completing downstream tasks on the private dataset through DP-SGD algorithm. On the one hand, due to the appearance of downstream tasks in the training phase, the trained model contains key information which plays an important role in completing downstream tasks. On the other hand, due to the usage of DP-SGD, regardless of how the trained model is used, no private information is disclosed. Thus, we utilize the trained model as an approximation of data distribution of the private dataset to synthesize datasets. Specifically, we generate data attributions of synthetic data record first, and then input these attributions into the trained model to gain a corresponding label. The generated data attributions and the gained labels are combined as one data record in the synthetic dataset. To make the synthetic dataset as similar as possible to the original private dataset, we generate data attributions of synthetic data records by randomly attribution-wise shuffling data attributions of original data records. By taking specific characteristics of downstream tasks into account, we successfully improve the data utility of synthetic datasets and our main contributions are as follows                     	
	\begin{itemize}
		
		\item{ \textbf{Superior Utility.} Our method improves the data utility of DP dataset synthesis significantly. By introducing downstream tasks into the training phase, our method successfully captures the key features which are important for improving performance of downstream tasks. The preservation of key features enhances the data utility of synthetic datasets in downstream tasks. For example, our method attains a maximum accuracy improvement of 2.40× and a maximum F1-score improvement of 15.28× according to experiment results.}
               
		\item{ \textbf{High Efficiency.} Our method can efficiently synthesize a dataset. In particular, our method utilizes the model trained on the private dataset as the sampler of original data distribution, and utilizes the inference procedure of trained model as the sample procedure. Our method can synthesize a dataset fast because the inference procedure of AI model is executed fast. Across various downstream tasks, our method attains a maximum efficiency improvement of 333.73× according to experiment results.}

        \item{ \textbf{Strong Scalability.} Our method has strong scalability than existing methods. As the data volume of original private dataset increases, existing methods such as PrivMRF \cite{cai2021privMRF} may fail to synthesize a dataset. Different from existing methods, our method trains an AI model on private datasets and utilizes the trained AI model to synthesize data records. As the data volume of original private dataset increases, the training procedure of AI model still can be executed although the training time may increase a little. Thus, as the data volume increases, our method does not fail. That is, our method has strong scalability. Across various downstream tasks, our method attains a maximum accuracy improvement of 1.59× when the dataset scale is doubled, and a maximum accuracy improvement of 1.62× when the dataset scale is tripled.}
		
	\end{itemize}
	
	
	\section{Related works}
    
    
    \par In the field of DP data synthesis, existing works are primarily divided into two categories: select-measure-generate methods and GAN-based methods.

    \par The select-measure-generate method leverages statistical correlation to synthesize datasets. In the selection phase, a set of queries such as low-dimensional distributions are selected to guarantee that important information is preserved. In the measure phase, answers to selected queries are computed in a differentially private way such that private information is not disclosed. In the phase of generation, a synthetic dataset is generated such that the selected queries have the same answers on the synthetic dataset as on the original private dataset. There are many works of select-measure-generate such as  \cite{jia2024absyn} \cite{zhang2021privsyn} \cite{cai2021privMRF} \cite{mckenna2022aim} \cite{mckenna2021MST} \cite{hardt2012mwem}. While the select-measure-generate method can preserve correlation among data attributions well, it suffers from severe scalability issues. When a private dataset is large, the select-measure-generate method may fail.


    \begin{figure*}[h]
		\centering
		\includegraphics[width=0.89\textwidth]{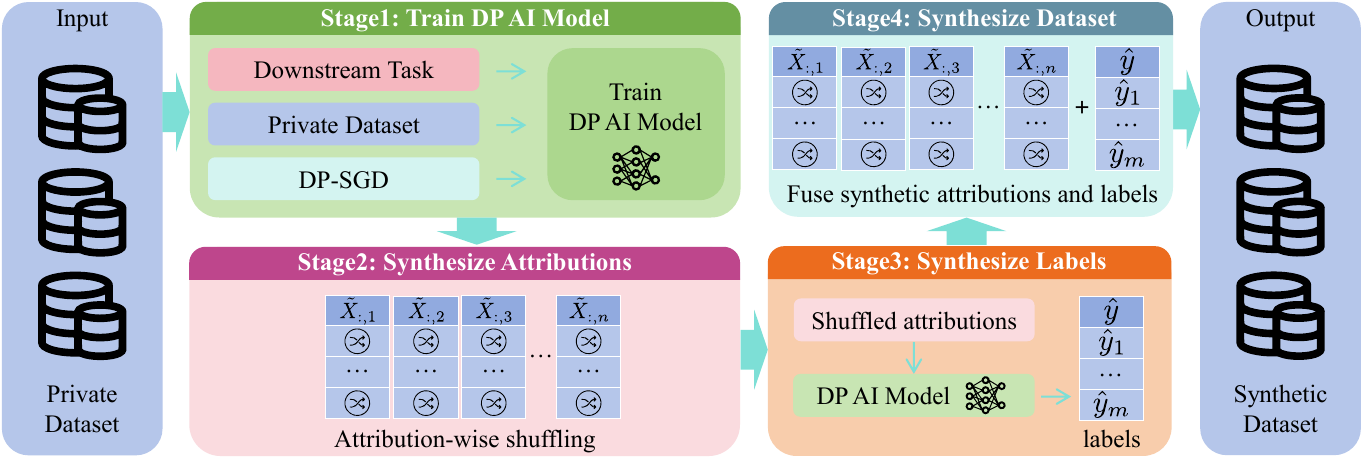}
		\caption{The workflow of DPDSyn.}
		\label{fig:workflow}
	\end{figure*}        

    \par GAN-based methods use adversarial learning to synthesize datasets. In particular, GAN-based methods train a generator for synthesizing datasets and a discriminator for evaluating similarity between synthesized dataset and original private dataset. The training phase of discriminator uses DP-SGD so that private information is preserved. There are many GAN-based methods such as \cite{srivastava2019GAN} \cite{ma2023rdp} \cite{hou2023wdp}. The GAN-based methods offer higher efficiency but often lack rigorous quality guarantees for data utility. The utility degradation in GAN-based methods primarily stems from the instability of adversarial training when the gradient is perturbed by differentially private noise.   

	\section{Method}
     \par In this section, our method DPDSyn is presented in detail. Specifically, our idea is elaborated first. Then, the workflow of our method is presented through the detailed description of each step. Finally, the privacy analysis of our method is given concisely.

	\subsection{Idea}    

    \par Our basic idea is to preserve as much information as possible that is critical to improving the performance of target downstream tasks, regardless of other information when synthesizing datasets. In general, a dataset contains much information. However, only partial information contained in the dataset is useful for a particular downstream task. The rest information is useless for the particular downstream task although the rest information may be critical for other downstream tasks. Our basic idea is to preserve useful information for target downstream tasks more when synthesizing datasets so that the data utility of synthetic datasets could be improved further. 

    \par Our idea is different from the idea of existing methods of DP dataset synthesis. In particular, the most existing methods aim at synthesizing datasets such that the distribution of synthetic dataset is as similar as possible to the original dataset distribution. For example, ABSyn \cite{jia2024absyn} synthesizes datasets such that selected low-dimensional distributions of synthetic dataset are the same as low-dimensional distributions of original dataset. The main idea of ABSyn is that dataset's distributions could be similar with a high probability when low-dimension distributions of these datasets are the same. The existing methods such as ABSyn do not pay enough attention to really useful information for downstream tasks, which leaves us the chance to improve the data utility of synthetic dataset further. 

    \par To achieve our idea, our method introduces downstream tasks into the procedure of synthesizing datasets to preserve critical information for downstream tasks. In particular, our method trains a DP AI model which can complete the downstream task on original private dataset through DP-SGD algorithm. The trained model contains the critical information needed by downstream tasks due to the appearance of downstream tasks in the training phase and does not disclose the private information of original dataset due to the usage of DP-SGD algorithm. Our method utilizes the trained DP AI model to generate synthetic datasets such that useful information for downstream tasks could be preserved as much as possible, resulting in that the data utility of synthetic dataset could be improved significantly.

	\subsection{Workflow}
    \par As shown in Figure \ref{fig:workflow}, there are four stages in our method workflow. Specifically, in the first stage, to preserve as much critical information as possible for improving the performance of downstream tasks, our method trains a task-aware DP AI model via DP-SGD algorithm to privately capture the mapping from data attributions to labels. In the second stage, to keep the distribution of each attribution unchanged in synthetic datasets, our method only shuffles each attribution randomly and then lets the shuffled attributions be new attributions of synthetic data records. In the third stage, the shuffled attributions are fed into the trained DP AI model to obtain the corresponding label so that the procedure of synthesizing data record can be completed efficiently. In the fourth stage, the shuffled attributions and the obtained labels are merged to form the final synthetic dataset. Next, each stage in workflow of our DPDSyn is elaborated on. 

    \par In the first stage, our method utilizes the private dataset to train a DP AI model which can complete the target downstream task. The trained DP serves two purposes. The first purpose is to capture critical information (which can enable the completion of downstream tasks) so that the data utility of synthetic dataset is improved significantly. The second purpose is to avoid disclosing private information by applying DP mechanisms in the training phase. That is, the first stage achieves the goal of improving data utility while preserving privacy.     

    \par To achieve purposes of the first stage, our method identifies the target downstream task and determines the corresponding model architecture. Specifically, our method first initializes training set size $N$ and establishes reasonable hyperparameters, such as batch size $b$, the number of epochs $E$, and learning rate $\eta$. To guarantee good performance of trained DP AI model, our method turns the minibatch size $\mathcal{B}_t$ and clipping bound $\mathcal{C}$ using a DP-style warm-up procedure. To guarantee that the final DP AI model satisfies ($\varepsilon$, $\delta$)-DP, our method computes the noise multiplier $z$ using the \textit{compute$\_$noise} function provided by the tensorflow-privacy library \footnote{\url{https://github.com/tensorflow/privacy}}. This function can incorporate the privacy parameters $(\varepsilon,\delta)$ with DP-SGD training configuration obtained from the warm-up stage, i.e., $(N, b, E)$, producing the required Gaussian noise scale for gradient perturbation.

    \par The training procedure of DP AI model is outlined in Algorithm \ref{alg:dp-train}. In particular, Algorithm \ref{alg:dp-train} takes the private dataset $\mathcal{D}$ and privacy parameters ($\varepsilon$, $\delta$) as inputs. Steps 1–5 prepare hyperparameter optimization to determine the training hyperparameters. During steps 1–4, the size of training set $\mathcal{D}_{train}$ is initialized. Then, by performance tuning and sensitivity calibration, necessary parameters are set, including the optimal batch size $b$, number of epochs $E$, learning rate $\eta$, the minibatch size $\mathcal{B}_t$, and clipping bound $\mathcal{C}$. In step 5, the noise multiplier $z$ is computed to ensure that the resulting DP AI model satisfies ($\varepsilon$, $\delta$)-DP. With all hyperparameters determined, steps 6–17 perform the DP-SGD training procedure and finally output the trained DP AI model.

    \begin{algorithm}
		\caption{DP AI Model Training with DP-SGD}
		\label{alg:dp-train}
		\begin{algorithmic}[1]
			\renewcommand{\algorithmicrequire}{\textbf{Input:}}
			\renewcommand{\algorithmicensure}{\textbf{Output:}}
			
			\REQUIRE Private dataset $\mathcal{D}=\{(x_i,y_i)\}$, 
			privacy budget $(\varepsilon, \delta)$
			\ENSURE DP AI model $\mathcal{M}_{DP}$
			
			\textbf{Stage 1: Hyperparameter Optimization}
			\STATE Let training set $\mathcal{D}_{train} $ be $ 80\%$ data samples from $\mathcal{D}$
			\STATE Initialize $N \gets |\mathcal{D}_{train}|$
			\STATE Performance Tuning: Perform grid search on $\mathcal{D}_{train}$ to determine the optimal batch size $b$, number of epochs $E$, and learning rate $\eta$ for standard SGD convergence.
			\STATE Sensitivity Calibration: Calibrate minibatch $\mathcal{B}_{t}$ and the clipping bound $\mathcal{C}$.
            
			\textbf{ Stage 2: Compute noise multiplier}
			\STATE$z \gets compute$\_$noise(N, b, E, \varepsilon, \delta)$ 
            
			\textbf{ Stage 3: DP training}
			\STATE Initialize model parameters $\theta_0$
			\FOR{$t = 1$ to $E$}
			\STATE Randomly sample a minibatch $\mathcal{B}_t \subset \mathcal{D}_{train}$
			\FOR{each sample $(x_i, y_i) \in \mathcal{B}_t$}
			\STATE Compute gradient: $g_t(x_i) \gets \nabla_{\theta_t} \mathcal{L}(\theta_t; x_i, y_i)$
			\STATE Clip gradient:\\
            $ \tilde{g}_t(x_i) \gets g_t(x_i) / \max(1, \frac{\|g_t(x_i)\|_2}{\mathcal{C}})$
			\ENDFOR
			\STATE Add noise: 
			$\bar{g}_t \gets \frac{1}{|\mathcal{B}_t|} \sum_{i \in \mathcal{B}_t} \left( \tilde{g}_t(x_i) + \mathcal{N}(0, z^2 \mathcal{C}^2 \mathcal{I}) \right)$
			\STATE Update parameters: $\theta_{t+1} \gets \theta_t - \eta_t \bar{g}_t$
			\ENDFOR
			
			\STATE Build DP AI model with final parameters: $\mathcal{M}_{DP} \gets \theta_E$
			\STATE \textbf{return} $\mathcal{M}_{DP}$
		\end{algorithmic}
	\end{algorithm}

    \par In the second stage, our method synthesizes high utility and privacy-preserving attributions of synthetic dataset. In particular, our method removes original labels $Y$ from the private dataset $\mathcal{D}$ and shuffles remaining attributions denoted by $X$ one by one, where the shuffling operation is denoted by $f$. The shuffled attribution (denoted by $\tilde{X}$) preserves the marginal distribution of each attribution while breaking data attribution correlation and data record correlation. Preserving the marginal distribution of each attribution makes the data attribution of synthetic dataset similar to that of original private dataset. Thus, the information contained in the synthetic dataset could be similar to the information contained in the original private dataset. In other words, the data utility of synthetic dataset could be similar to the data utility of original private dataset. Breaking data attribution correlation and data record correlation makes it impossible to infer information about original data record in private dataset by synthetic dataset, thus preserving the private information contained in the private dataset. In addition, the shuffling procedure can be performed efficiently so this procedure allows our method to rapidly obtain the new attribution for synthetic dataset. In a word, by shuffling, our method can efficiently synthesize privacy-preserving attributions with high data utility for synthetic datasets.  

    \par In the third stage, our method synthesizes labels for synthetic attributions by DP AI model obtained in the first stage. In particular, our method feeds shuffled attributions $\tilde{X}$ into the DP AI model $\mathcal{M}_{DP}$ to obtain corresponding labels $\hat{Y}$. Our method successfully transforms useful information for downstream tasks from the DP AI model to synthetic datasets by the procedure of synthesizing labels. Different from the traditional select-measure-generate method that relies on iterative marginal selection, our method leverages the DP AI model to efficiently synthesize data records. Our method bypasses the heavy computational burden associated with iterative selection of low-dimensional distribution and generation of high-dimensional data record by low-dimensional distribution, significantly reducing both time and resource consumption.

    \par In the fourth stage, our method fuses the synthetic attributions $\tilde{X}$ with synthetic labels $\hat{Y}$ to output the synthetic dataset. The way in which our method synthesizes datasets is different from the way of GAN-based method although our method and GAN-based method both train an AI model. In particular, in the training phase of GAN-based method, the downstream task is not taken into account, resulting in that the trained model in GAN-based methods does not focus enough on the information critical for completing downstream tasks. On the contrary, our method introduces downstream tasks into the training phase of DP AI model, which enables the model to retain a lot of useful information for completing downstream tasks. 
    
	\par The workflow of DPDSyn is formally presented in Algorithm \ref{alg:dataset-gen-corrected}. Given the private dataset $\mathcal{D}$ and privacy parameters ($\varepsilon$, $\delta$), steps 1-2 identify the downstream task and specify a corresponding AI model, respectively. In step 3, Algorithm \ref{alg:dp-train} is employed to obtain a DP AI model. In steps 4–8, synthetic attributions $\tilde{X}$ are obtained by applying column-wise shuffling to the attributions $X$ from original dataset $\mathcal{D}$ after labels $Y$ are removed. In step 9, the synthetic attributions $\tilde{X}$ are fed into the DP AI model $\mathcal{M}_{DP}$ to synthesize corresponding labels $\hat{Y}$. Finally, synthesized labels $\hat{Y}$ and synthesized attributions $\tilde{X}$ are combined to produce the final privacy-preserving synthetic dataset $\hat{\mathcal{D}}$.
		
	\begin{algorithm}
		\caption{DPDSyn}
		\label{alg:dataset-gen-corrected}
		\begin{algorithmic}[1]
			\renewcommand{\algorithmicrequire}{\textbf{Input:}}
			\renewcommand{\algorithmicensure}{\textbf{Output:}}
			
			\REQUIRE Private dataset $\mathcal{D}$, privacy parameters $(\varepsilon, \delta)$
			\ENSURE Synthetic dataset $\hat{\mathcal{D}}$ 
			
            \textbf{Stage 1: Train DP AI Model}
			\STATE Identify downstream task
            \STATE Identify an AI model suitable for the downstream task
			\STATE Train DP AI model using Algorithm \ref{alg:dp-train}:\\
			$\mathcal{M}_{DP} \gets \text{DP\_Train}(\mathcal{D}, \varepsilon, \delta)$
			
            \textbf{Stage 2: Synthesize Attributions}
            \STATE Extract attribution matrix $X \gets \mathcal{D} \setminus Y$,\\
                  where $Y$ denotes data labels.            
			\STATE Initialize shuffled attribution matrix $\tilde{X}$
			
			\FOR{each attribution column in $X$}
			\STATE $\tilde{X}_{:,j} \gets f(X_{:,j})$, where $f$ represents randomly shuffling operation ~~~~~~~{\small$\triangleright$ column-wise permute $j$-th attribution}\\
            
			\ENDFOR
			
            \textbf{Stage 3: Synthesize Labels}
			\STATE Synthetic labels $\hat{Y} \gets \mathcal{M}_{DP}(\tilde{X})$ 
			 
            \textbf{Stage 4: Synthesize Dataset}
            \STATE Combine synthetic attributions and labels:\\ 
			$\hat{\mathcal{D}} \gets \{\tilde{X}, \ \hat{Y}\}$
			\STATE \textbf{return} $\hat{\mathcal{D}}$
		\end{algorithmic}
	\end{algorithm}
	
	\subsection{Analysis of Privacy Guarantee}
    \par Our method satisfies $(\varepsilon, \delta)$ differential privacy. In particular, our method uses the private dataset only in the first and the second stage in Algorithm \ref{alg:dataset-gen-corrected}. In the first stage, our method uses DP-SGD algorithm to train a DP AI model on the private dataset through $(\varepsilon, \delta)$ DP mechanism. Thus, our method satisfies $(\varepsilon, \delta)$ differential privacy. In the second stage, our method removes labels of data records and randomly shuffles attributions to break data attribution correlation and data record correlation so that there are no private information disclosed. In a word, our method satisfies $(\varepsilon, \delta)$ differential privacy according to the Sequential Composition theorem \cite{dwork2014algorithmic}.
	
	\section{Experiments}
    \par In this section, comprehensive experiments are conducted to evaluate our method DPDSyn and eight existing state-of-the-art methods in terms of data utility, efficiency, and scalability. Specifically, the first experiment evaluates the data utility of synthetic dataset by measuring the performance of downstream tasks. The second experiment evaluates efficiency by runtime as well as the trade-off between accuracy and runtime. The third experiment evaluates scalability by expanding the scale of private dataset to 2× and 3× its original scale. 
    
	\subsection{Setting}
	
    \subsubsection{Datasets}
    \par To comprehensively evaluate all methods, the evaluation experiments are conducted  on four widely used benchmark datasets including Adult \cite{kohavi1996scaling}, Br2000 \cite{IPUMS_USA_2015}, LPD \cite{LiverDiseasePatientDataset_Kaggle}, and Smoking \cite{Smoking_Signal_of_Body_Classification}. The information of these datasets is summarized in Table \ref{tab1}, where “Cat. Feat.” represents the number of categorical attributions and “Num. Feat.” represents the number of numerical attributes. 

	\begin{table}[h!]
		\centering
		\caption{Summary of datasets used in experiments}
		\label{tab1}
		\resizebox{\columnwidth}{!}{
			\begin{tabular}{cccccp{4.2cm}}
				\toprule 
				Dataset & Record   & Attribution   & Cat. Feat. & Num. Feat. & Task \\ \midrule
				Adult    & 48842 & 14       & 7 & 6 & Whether an individual’s \mbox{income} exceeds \$50K \\
				Br2000 & 38000 & 14         & 12 & 1 & Whether an individual \mbox{is} in the high-income group   \\
				LPD   & 27158 & 11  & 1  & 9 & Whether an individual has liver disease  \\
				Smoking & 55692 & 25 & 6 & 18& Whether an individual is a smoker   \\ \bottomrule
			\end{tabular}
		}
	\end{table}

    \par There are three points worth mentioning. First, in Smoking dataset, the attribute \textit{oral} is removed because its value is constant across all data records. Second, in LPD dataset, the samples containing null values are removed, and float values are transformed into integers so that select-measure-generate methods can apply. Third, the preprocessed datasets are randomly split into training, validation, and test sets with a ratio of 8:1:1. 

      
    \subsubsection{Baselines} Eight state-of-the-art baselines are used in comparison experiments, including ABSyn \cite{jia2024absyn}, PrivSyn \cite{zhang2021privsyn}, PrivPetal \cite{cai2025privpetal}, PrivMRF \cite{cai2021privMRF}, AIM \cite{mckenna2022aim}, MST \cite{mckenna2021MST}, MWEM+PGM \cite{hardt2012mwem} and DP-GAN \cite{srivastava2019GAN}. The DP-GAN uses GAN to synthesize datasets. Other methods are all the select-measure-generate method while the generation method within these methods varies. Specifically, PrivPetal and PrivMRF use MRF to generate synthetic datasets. ABSyn, AIM and MST use PGM to generate synthetic datasets. Following the setting of ABSyn, the generation method MW of MWEM is replaced by Private-PGM in this paper to accelerate the generation procedure. PrivSyn uses GUM to generate synthetic datasets.
        
    \subsubsection{Downstream Models and Tasks } To evaluate all dataset synthesis methods, three representative AI models are used to complete the downstream task, namely MLP, SVM, and FT-Transformer. In particular, MLP and SVM are representative models of classical machine learning  and FT-Transformer is a representative model of modern deep learning. These representative downstream models make the evaluation more comprehensive. The downstream tasks vary as described in Table \ref{tab1}.

    \subsubsection{Evaluation Way}
    \par Each dataset synthesis method utilizes the private dataset to synthesize a dataset and an AI model for completing downstream task is trained on the synthesized dataset. Then, the trained AI model is subsequently evaluated on the original private test set to obtain performance metrics, including accuracy, F1-score, and runtime. To ensure statistical stability of experiment results, all experiments are executed five times and reported results are the average result of the five experiments. Notably, the following dataset–method pairs are excluded from the evaluation due to experiment runtime exceeding 6 hours: Smoking dataset under PrivSyn; Smoking and LPD datasets under PrivMRF.

	\subsection{Evaluation}

 	\subsubsection{Utility}
        
    \par In the first experiment, the data utility of synthetic dataset is evaluated by accuracy and F1-score of downstream models including MLP, SVM, and FT-Transformer. The experiment results are shown in Table \ref{tab:utility1}, \ref{tab:utility2}, \ref{tab:utility3}, and \ref{tab:utility4}. In these tables, NP Baseline denotes results on the private dataset without differentially private protection, serving as the empirical performance upper bound. Max. Imp. and Avg. Imp. denote the maximum and average performance improvements, respectively. $T/O$ denotes timeout. $0.00^\dagger$ indicates utility collapse where the model fails to predict the minority class. 
    

    \begin{table}[thb!]
		\centering
        \caption{Utility Comparison of Adult Dataset.}
		\label{tab:utility1}
        \setlength{\tabcolsep}{3pt} 
        \footnotesize 
        \resizebox{\columnwidth}{!}{
		      \begin{tabular}{ccccccc}
			\toprule
			\multirow{2}{*}{Methods}     & \multicolumn{2}{c}{MLP}             & \multicolumn{2}{c}{SVM}             & \multicolumn{2}{c}{FT-Transformer}     \\ \cline{2-7}
			& \rule{0pt}{2.6ex}            acc              & F1-score         & acc              & F1-score         & acc              & F1-score         \\ \midrule
            \textbf{NP Baseline}         & \textbf{0.8528}           & \textbf{0.6410}           & \textbf{0.8452}           & \textbf{0.6131}           & \textbf{0.8592}           & \textbf{0.6587}           \\ \midrule
			ABSyn                        & 0.8352           & 0.5663           & 0.8347           & 0.5698           & 0.8386           & 0.6257           \\
			PrivSyn                      & 0.7426           & 0.0888           & 0.7653           & 0.0392           & 0.6959           & 0.1355           \\
			PrivMRF                      & 0.8342           & 0.5814           & 0.8331           & 0.5735           & 0.8366           & 0.6288           \\
			PrivPetal                    & 0.8237           & 0.5511           & 0.8228           & 0.5090           & 0.8291           & 0.6056           \\
			AIM                          & 0.8368           & 0.6011           & 0.8351           & 0.5807           & 0.8379           & 0.6187            \\
			MST                          & 0.7845           & 0.1692           & 0.7993           & 0.2686           & 0.7771           & 0.1623           \\
			MWEM                         & 0.7893            & 0.2330          & 0.7918           & 0.2356           & 0.7832           & 0.2676           \\
			DP\_GAN                      & 0.5409            & 0.2320          & 0.5306           & 0.2672           & 0.5604           & 0.2449           \\
			\textbf{DPDSyn}               & \textbf{0.8429}  & \textbf{0.6260}  & \textbf{0.8398}  & \textbf{0.5991}  & \textbf{0.8429}  & \textbf{0.6288}  \\ \midrule
			
			\textbf{Max. Imp.} & \textbf{1.56×} $\uparrow$  & \textbf{7.05×} $\uparrow$  & \textbf{1.58×} $\uparrow$  & \textbf{15.28×} $\uparrow$  & \textbf{1.50×} $\uparrow$  & \textbf{4.64×} $\uparrow$  \\ \midrule
			\textbf{Avg. Imp.} & \textbf{1.11×} $\uparrow$  & \textbf{2.56×} $\uparrow$  & \textbf{1.10×} $\uparrow$   & \textbf{3.33×} $\uparrow$  & \textbf{1.11×} $\uparrow$   & \textbf{2.19×} $\uparrow$ \\ \bottomrule
			
		\end{tabular}
        }
	\end{table}

    \begin{table}[thb!]
		\centering
        \caption{ Utility Comparison of Br2000 Dataset. }
		\label{tab:utility2}
        \setlength{\tabcolsep}{3pt} 
        \footnotesize 
        \resizebox{\columnwidth}{!}{
		      \begin{tabular}{ccccccc}
			\toprule
			\multirow{2}{*}{Methods}     & \multicolumn{2}{c}{MLP}             & \multicolumn{2}{c}{SVM}             & \multicolumn{2}{c}{FT-Transformer}     \\ \cline{2-7}
			& \rule{0pt}{2.6ex}            acc              & F1-score         & acc              & F1-score         & acc              & F1-score         \\ \midrule
            \textbf{NP Baseline}         & \textbf{0.8221}           & \textbf{0.6168}           & \textbf{0.8174}           & \textbf{0.6105}           & \textbf{0.8134}           & \textbf{0.6290}           \\ \midrule
			ABSyn                        & 0.6976           & 0.3148           & 0.8152           & 0.5997           & 0.8152           & 0.6113           \\
			PrivSyn                      & 0.7067           & 0.3197           & 0.7592           & 0.3168           & 0.6264           & 0.2660           \\
			PrivMRF                      & 0.8172           & 0.6197           & 0.8150           & 0.6053           & 0.8156           & 0.6237           \\
			PrivPetal                    & 0.7997           & 0.5746           & 0.8111           & 0.5874           & 0.7913           & 0.5754           \\
			AIM                          & 0.8158           & 0.6071           & 0.8154           & 0.6057           & 0.8169           & 0.6193            \\
			MST                          & 0.7882           & 0.5340           & 0.7930           & 0.5594           & 0.7908           & 0.5545           \\
			MWEM                         & 0.7917           & 0.5517           & 0.7937           & 0.5611           & 0.7926           & 0.5571           \\
			DP\_GAN                      & 0.5031           & 0.2723           & 0.5031           & 0.3064           & 0.5129           & 0.2883           \\
			\textbf{DPDSyn}               & \textbf{0.8158}  & \textbf{0.6087}  & \textbf{0.8189}  & \textbf{0.6241}  & \textbf{0.8105}  & \textbf{0.5853}  \\ \midrule
			
			\textbf{Max. Imp.} & \textbf{1.62×} $\uparrow$  & \textbf{2.24×} $\uparrow$  & \textbf{1.63×} $\uparrow$  & \textbf{2.04×} $\uparrow$  & \textbf{1.58×} $\uparrow$  & \textbf{2.20×} $\uparrow$  \\ \midrule
			\textbf{Avg. Imp.} & \textbf{1.13×} $\uparrow$  & \textbf{1.42×} $\uparrow$  & \textbf{1.10×} $\uparrow$   & \textbf{1.30×} $\uparrow$  & \textbf{1.12×} $\uparrow$   & \textbf{1.27×} $\uparrow$ \\ \bottomrule
			
		\end{tabular}
        }
	\end{table}

    \begin{table}[thb!]
		\centering
        \caption{Utility Comparison of Smoking Dataset.} 
		\label{tab:utility3}
        \setlength{\tabcolsep}{3pt} 
        \footnotesize 
        \resizebox{\columnwidth}{!}{
		      \begin{tabular}{ccccccc}
			\toprule
			\multirow{2}{*}{Methods}     & \multicolumn{2}{c}{MLP}             & \multicolumn{2}{c}{SVM}             & \multicolumn{2}{c}{FT-Transformer}     \\ \cline{2-7}
			& \rule{0pt}{2.6ex}            acc              & F1-score         & acc              & F1-score         & acc              & F1-score         \\ \midrule
            \textbf{NP Baseline}         & \textbf{0.7463}           & \textbf{0.6902}           & \textbf{0.7357}           & \textbf{0.6650}           & \textbf{0.7350}           & \textbf{0.6826}           \\ \midrule
			ABSyn                        & 0.6029           & 0.2479           & 0.6867           & 0.3872           & 0.6901           & 0.3910           \\
			PrivSyn                      & $T/O$              & $T/O$              & $T/O$              & $T/O$              & $T/O$              & $T/O$           \\
			PrivMRF                      & $T/O$              & $T/O$              & $T/O$              & $T/O$              & $T/O$              & $T/O$           \\
			PrivPetal                    & 0.7073           & 0.6110           & 0.7095           & 0.6167           & 0.7059           & 0.6179           \\
			AIM                          & 0.6918           & 0.5501           & 0.6945           & 0.5611           & 0.6889           & 0.5649            \\
			MST                          & 0.6329           & $0.00^\dagger$   & 0.6328            & $0.00^\dagger$           & 0.6308           & $0.00^\dagger$           \\
			MWEM                         & 0.5930           & 0.2723           & 0.6174           & 0.3556           & 0.5949           & 0.2937           \\
			DP\_GAN                      & 0.5097           & 0.3723           & 0.5222           & 0.3969           & 0.5212           & 0.3904           \\
			\textbf{DPDSyn}               & \textbf{0.7275}  & \textbf{0.6498}  & \textbf{0.7361}  & \textbf{0.6555}  & \textbf{0.7368}  & \textbf{0.6628}  \\ \midrule
			
			\textbf{Max. Imp.} & \textbf{1.43×} $\uparrow$  & \textbf{2.62×} $\uparrow$  & \textbf{1.41×} $\uparrow$  & \textbf{1.84×} $\uparrow$  & \textbf{1.41×} $\uparrow$  & \textbf{2.26×} $\uparrow$  \\ \midrule
			\textbf{Avg. Imp.} & \textbf{1.17×} $\uparrow$  & \textbf{1.90×} $\uparrow$  & \textbf{1.14×} $\uparrow$   & \textbf{1.70×} $\uparrow$  & \textbf{1.15×} $\uparrow$   & \textbf{1.76×} $\uparrow$ \\ \bottomrule
			
		\end{tabular}
        }
	\end{table}

    \begin{table}[thb!]
		\centering
        \caption{Utility Comparison of LPD Dataset .}
		\label{tab:utility4}
        \setlength{\tabcolsep}{3pt} 
        \footnotesize 
        \resizebox{\columnwidth}{!}{
		      \begin{tabular}{ccccccc}
			\toprule
			\multirow{2}{*}{Methods}     & \multicolumn{2}{c}{MLP}             & \multicolumn{2}{c}{SVM}             & \multicolumn{2}{c}{FT-Transformer}     \\ \cline{2-7}
			& \rule{0pt}{2.6ex}            acc              & F1-score         & acc              & F1-score         & acc              & F1-score         \\ \midrule
            \textbf{NP Baseline}         & \textbf{0.7647}           & \textbf{0.8542}           & \textbf{0.7077}           & \textbf{0.8240}           & \textbf{0.8233}           & \textbf{0.8786}           \\ \midrule
			ABSyn                        & 0.6918           & 0.8108           & 0.7035           & 0.8217           & 0.7124           & 0.8219           \\
			PrivSyn                      & 0.3063           & 0.0727           & 0.2949           & $0.00^\dagger$   & 0.3362           & 0.2435           \\
			PrivMRF                      & $T/O$            & $T/O$            & $T/O$            & $T/O$            & $T/O$            & $T/O$           \\
			PrivPetal                    & 0.7029           & 0.8248           & 0.7025           & 0.8247           & 0.7079           & 0.8237           \\
			AIM                          & 0.7051           & 0.8269           & 0.7046           & 0.8265           & 0.7194           & 0.8366            \\
			MST                          & 0.7051           & 0.8270           & 0.7051           & 0.8270           & 0.7203           & 0.8374           \\
			MWEM                         & 0.6683           & 0.7552           & 0.6817           & 0.7997           & 0.6761           & 0.7610           \\
			DP\_GAN                      & 0.4905           & 0.4799           & 0.4923           & 0.5059           & 0.4915           & 0.4912           \\
			\textbf{DPDSyn}               & \textbf{0.7086}  & \textbf{0.8245}  & \textbf{0.7085}  & \textbf{0.8257}  & \textbf{0.7242}  & \textbf{0.8382}  \\ \midrule
			
			\textbf{Max. Imp.} & \textbf{2.31×} $\uparrow$  & \textbf{11.34×} $\uparrow$  & \textbf{2.40×} $\uparrow$  & \textbf{1.63×} $\uparrow$  & \textbf{2.15×} $\uparrow$  & \textbf{3.44×} $\uparrow$  \\ \midrule
			\textbf{Avg. Imp.} & \textbf{1.16×} $\uparrow$  & \textbf{1.26×} $\uparrow$  & \textbf{1.16×} $\uparrow$   & \textbf{1.25×} $\uparrow$  & \textbf{1.16×} $\uparrow$   & \textbf{1.22×} $\uparrow$ \\ \bottomrule
			
		\end{tabular}
        }
	\end{table}
    
    \par Based on experiment results in Table \ref{tab:utility1}, \ref{tab:utility2}, \ref{tab:utility3}, and \ref{tab:utility4}, the data utility of DPDSyn is consistently higher than that of other methods across different downstream models. In particular, on average, DPDSyn achieves a maximum accuracy improvement of 2.31× (0.7086/0.3063) in MLP, 2.40× in SVM, and 2.15× in FT-Transformer, compared to the eight benchmark baselines. Meanwhile, on average, DPDSyn achieves a maximum F1-score improvement of 11.34× in MLP, 15.28× in SVM, and 4.64× in FT-Transformer, compared to the eight benchmark baselines. The experimental results demonstrate that: the way in which our DPDSyn incorporates downstream task awareness into the synthesis procedure successfully preserves important task-relevant information, thereby effectively improving the data utility of synthetic dataset.

	\subsubsection{Efficiency} 

    \begin{figure*}[t]
		\centering
		
		\begin{subfigure}[b]{\textwidth}
			\centering
			\includegraphics[width=1.0\linewidth]{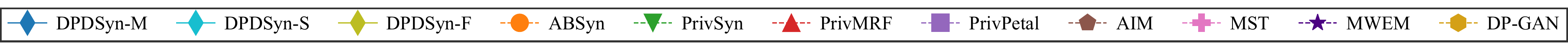}
			\label{fig:legend_for_time}
		\end{subfigure} 
		
		\vspace{-0.44cm}
		
		\begin{tabular}{c l}
            \multirow{4}{*}[6.5ex]{\rotatebox{90}{Time}}
			\setcounter{subfigure}{0}\renewcommand{\thesubfigure}{\alph{subfigure}}%
			\begin{minipage}[c]{0.92\linewidth}
				\centering
				\setlength{\tabcolsep}{0pt} 
				\renewcommand{\arraystretch}{0} 
				\setcounter{subfigure}{0}\renewcommand{\thesubfigure}{\alph{subfigure}}%
				\begin{subfigure}[b]{0.23\textwidth}
					\centering
					\includegraphics[width=\linewidth]{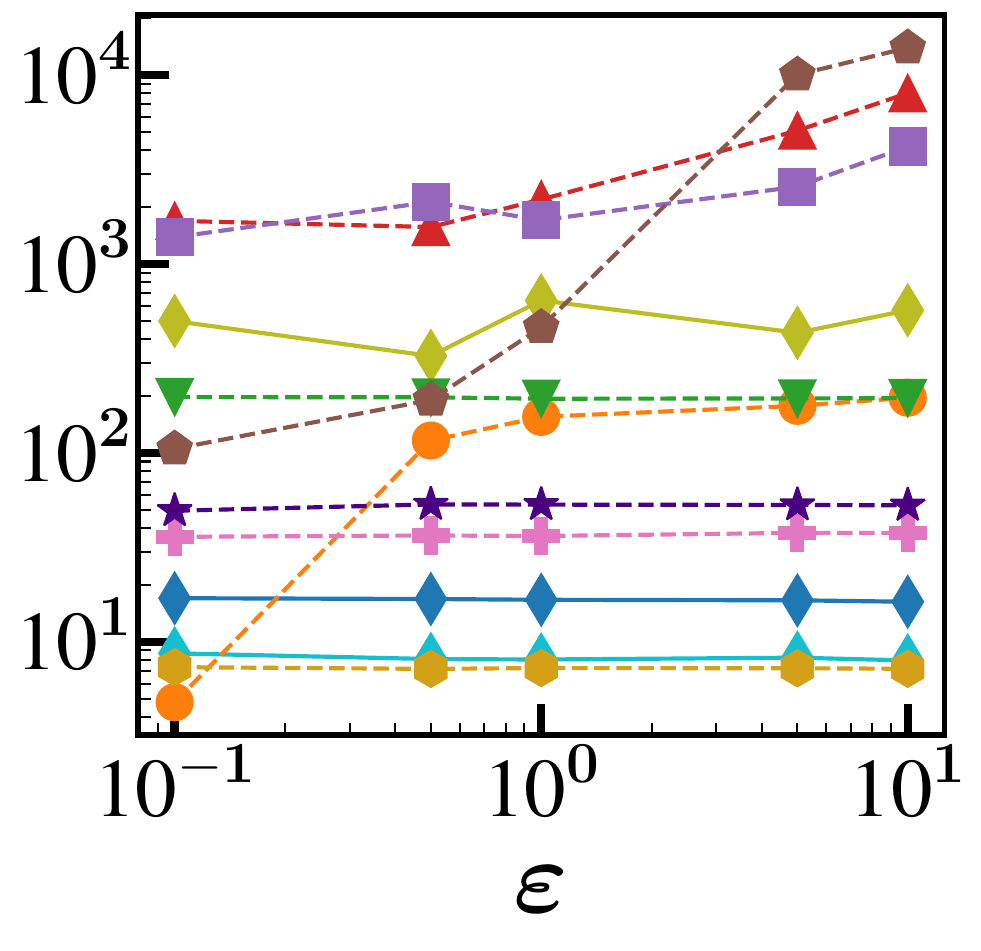}
					\caption{adult}
				\end{subfigure}\hspace*{-0.5mm}%
				\hfill
				\begin{subfigure}[b]{0.23\textwidth}
					\centering 
					\includegraphics[width=\linewidth]{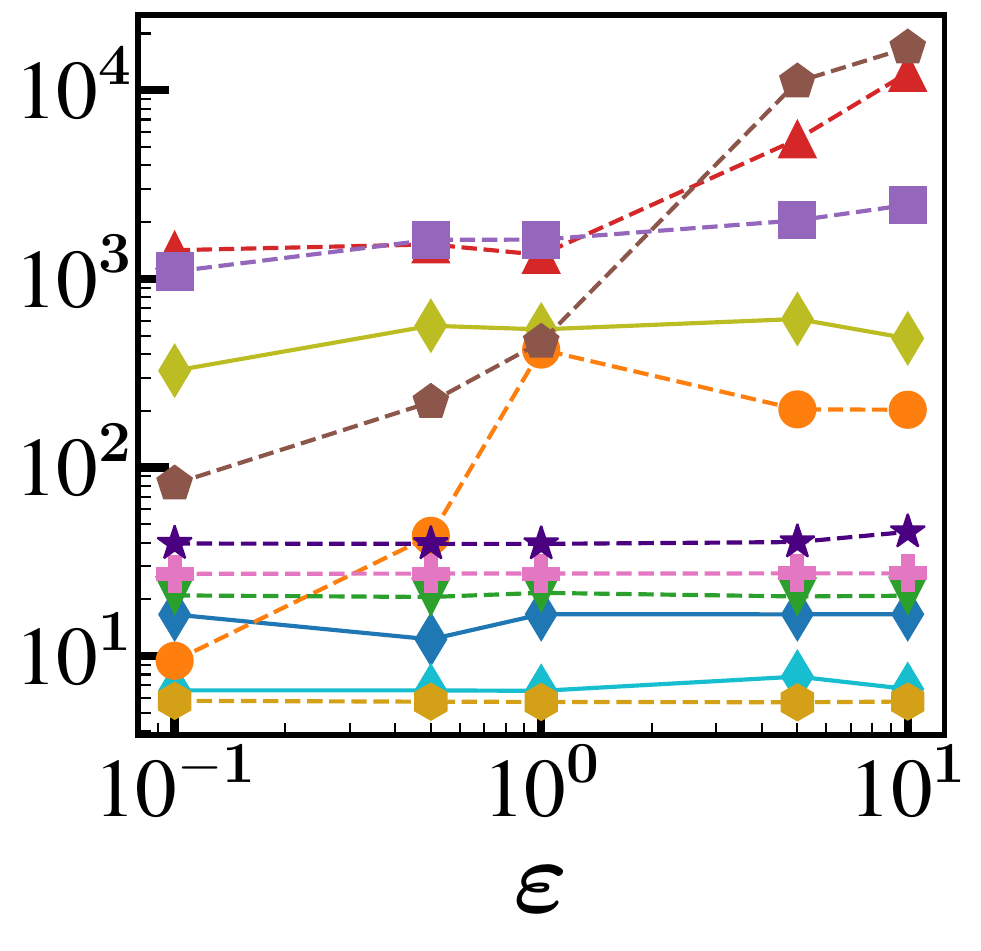}
					\caption{br2000}
				\end{subfigure}\hspace*{-0.5mm}%
				\hfill
				\begin{subfigure}[b]{0.23\textwidth}
					\centering
					\includegraphics[width=\linewidth]{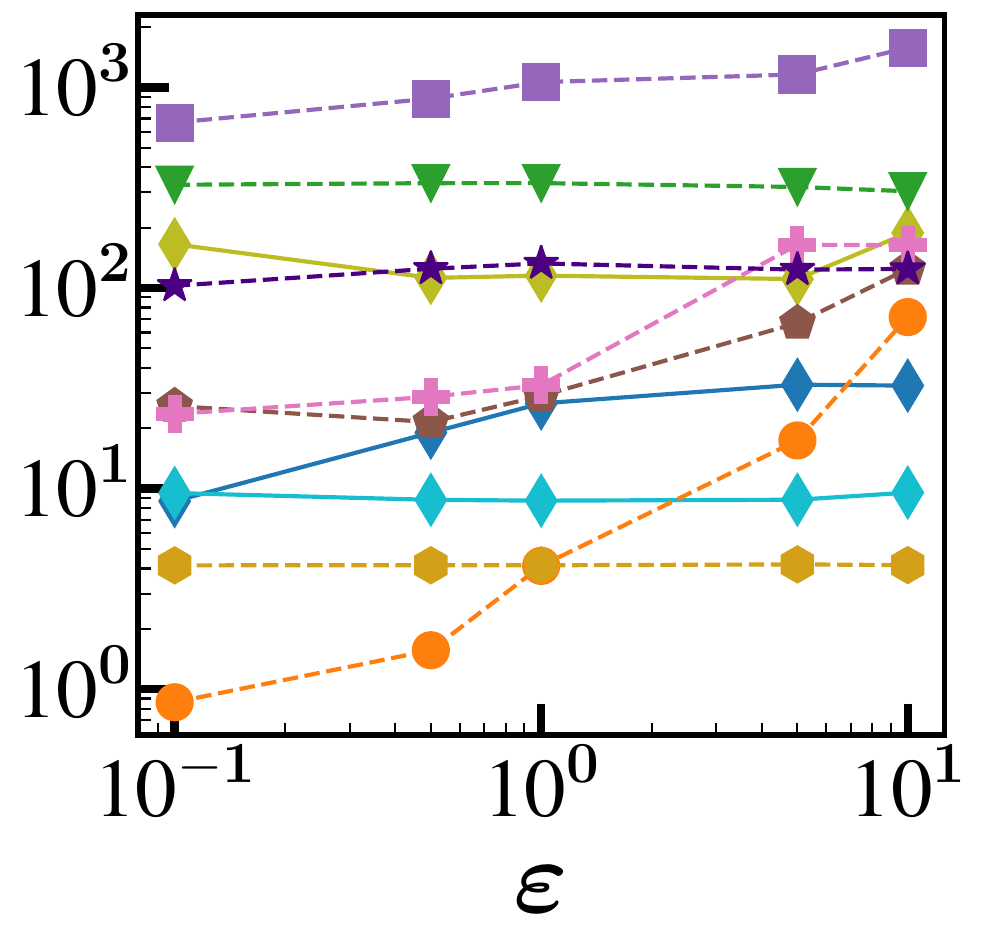}
					\caption{LPD}
				\end{subfigure}\hspace*{-0.5mm}%
				\hfill
				\begin{subfigure}[b]{0.23\textwidth}
					\centering
					\includegraphics[width=\linewidth]{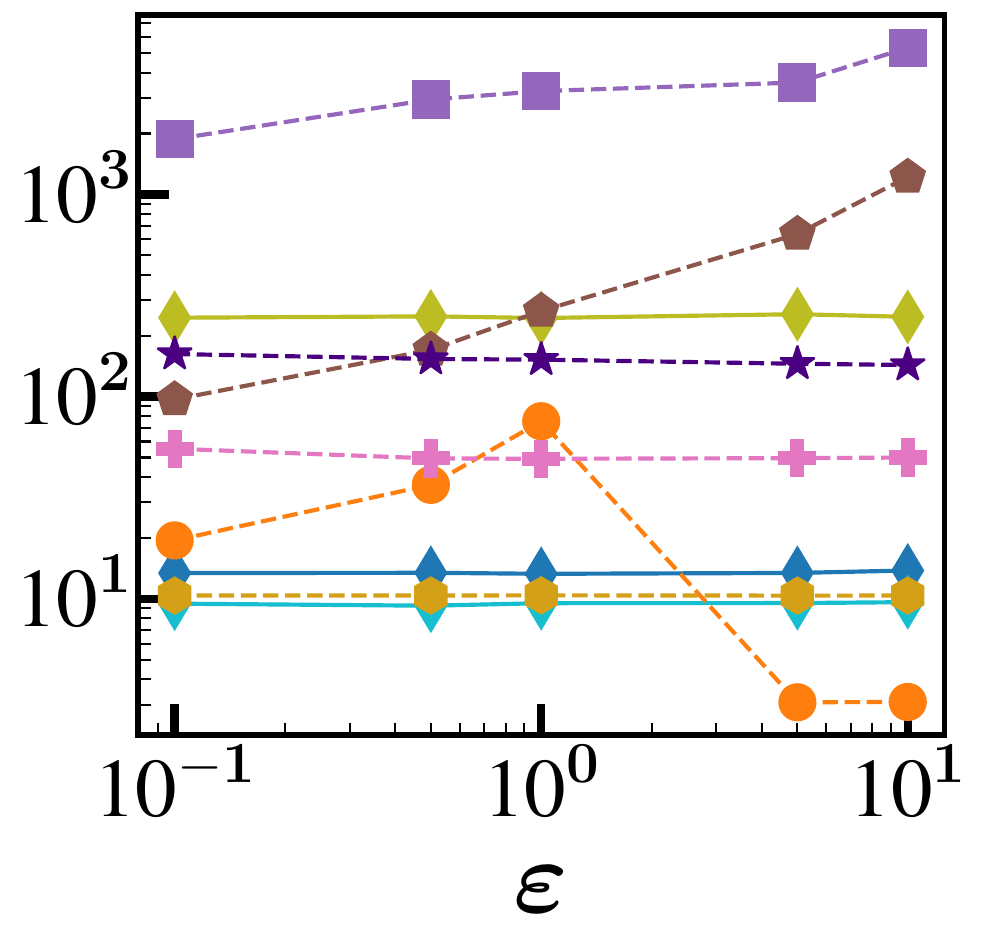}
					\caption{smoking}
				\end{subfigure}\hspace*{-0.5mm}%
			\end{minipage}
		\end{tabular}

		\caption{Runtime comparison. DPDSyn-\{M, S, F\} denote the runtime of DPDSyn under MLP, SVM, and FT-Transformer downstream models, respectively.  All other runtime of baseline method are independent of the downstream model.}

		\label{fig:time_overview_1x}
	\end{figure*} 
    
	\begin{figure*}[t]

		\centering
		
		\begin{subfigure}[b]{\textwidth}
			\centering
			\includegraphics[width=1.0\linewidth]{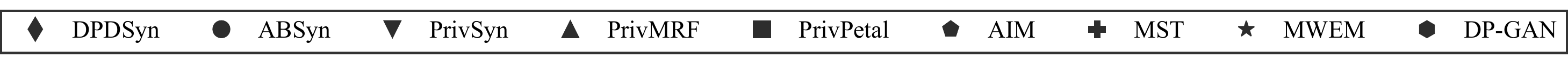}
		\end{subfigure} 
		
		\vspace{-0.12cm}
		
        \begin{tabular}{c l}
            \multirow{4}{*}[6.5ex]{\rotatebox{90}{Acc}} 
			\setcounter{subfigure}{0}\renewcommand{\thesubfigure}{\alph{subfigure}}%
			\begin{minipage}[c]{0.92\linewidth}
				\centering
				\setlength{\tabcolsep}{0pt} 
				\renewcommand{\arraystretch}{0} 
				\setcounter{subfigure}{0}\renewcommand{\thesubfigure}{\alph{subfigure}}%
				
				\begin{subfigure}[b]{0.23\textwidth}
					\centering
					\includegraphics[width=\linewidth]{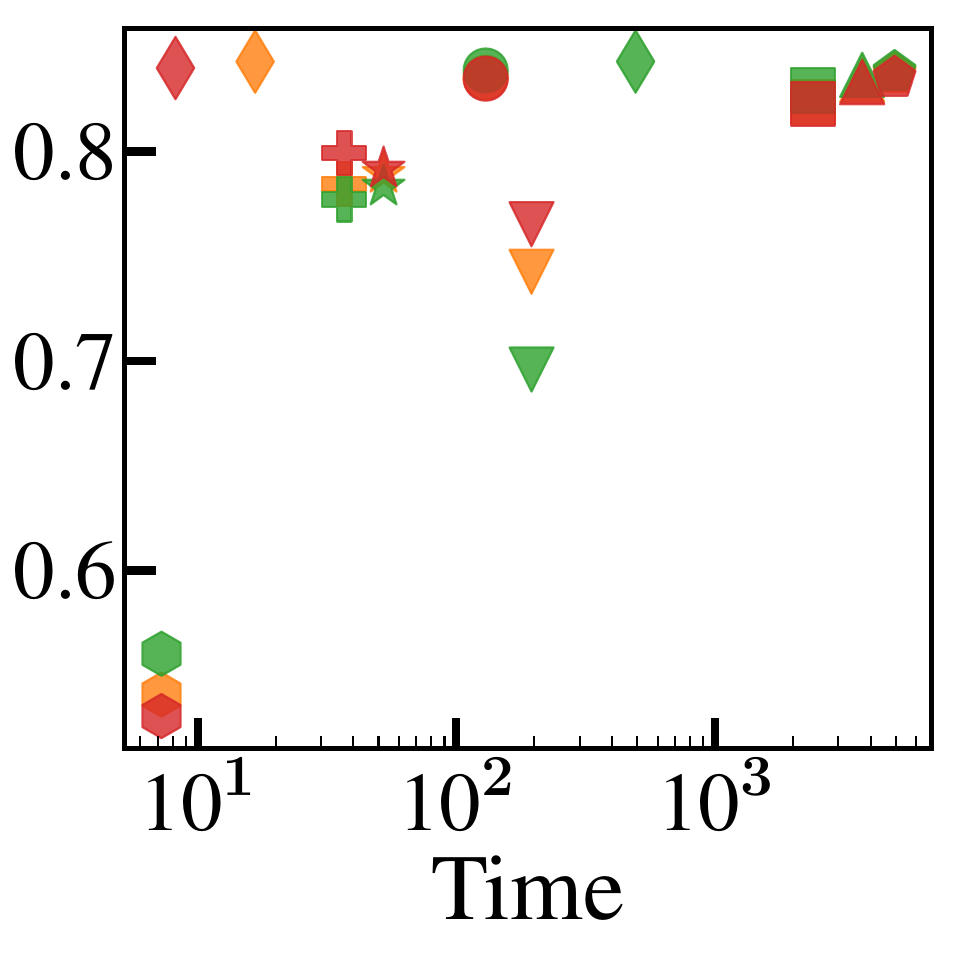}
					\caption{adult}
				\end{subfigure}\hspace*{-0.5mm}%
				\hfill
				\begin{subfigure}[b]{0.23\textwidth}
					\centering 
					\includegraphics[width=\linewidth]{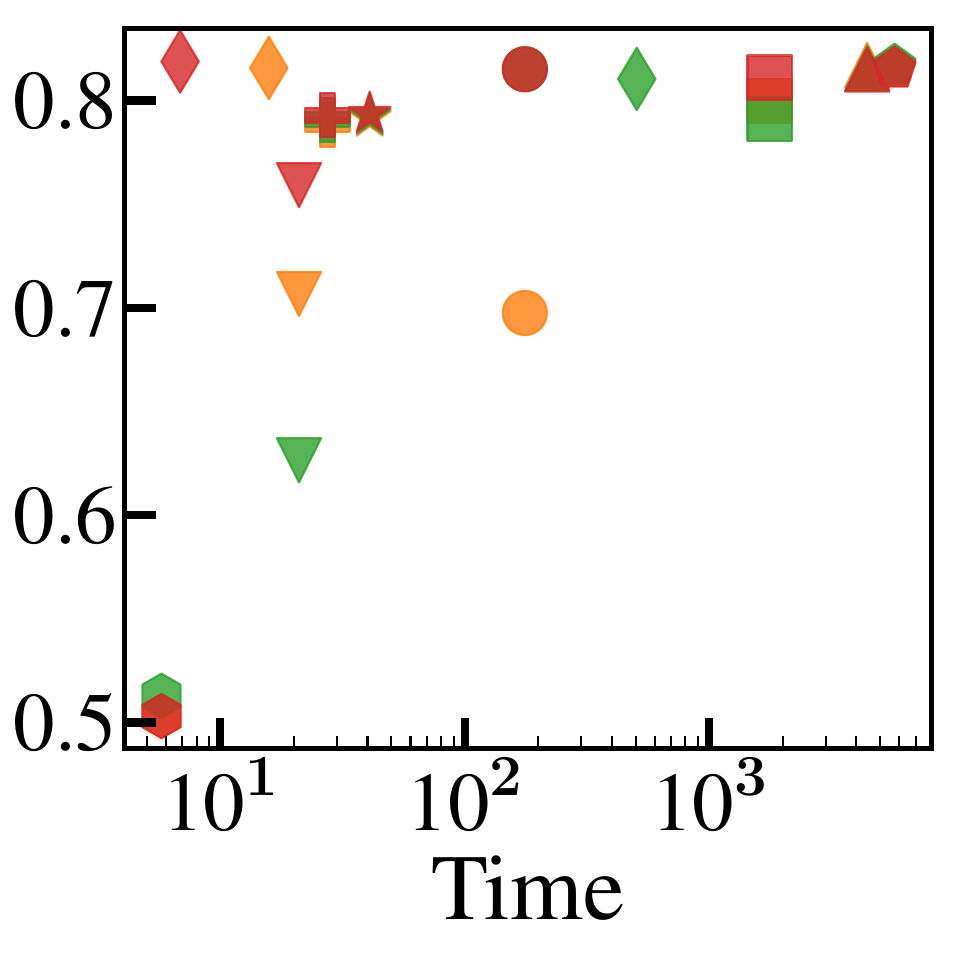}
					\caption{br2000}
				\end{subfigure}\hspace*{-0.5mm}%
				\hfill
				\begin{subfigure}[b]{0.23\textwidth}
					\centering
					\includegraphics[width=\linewidth]{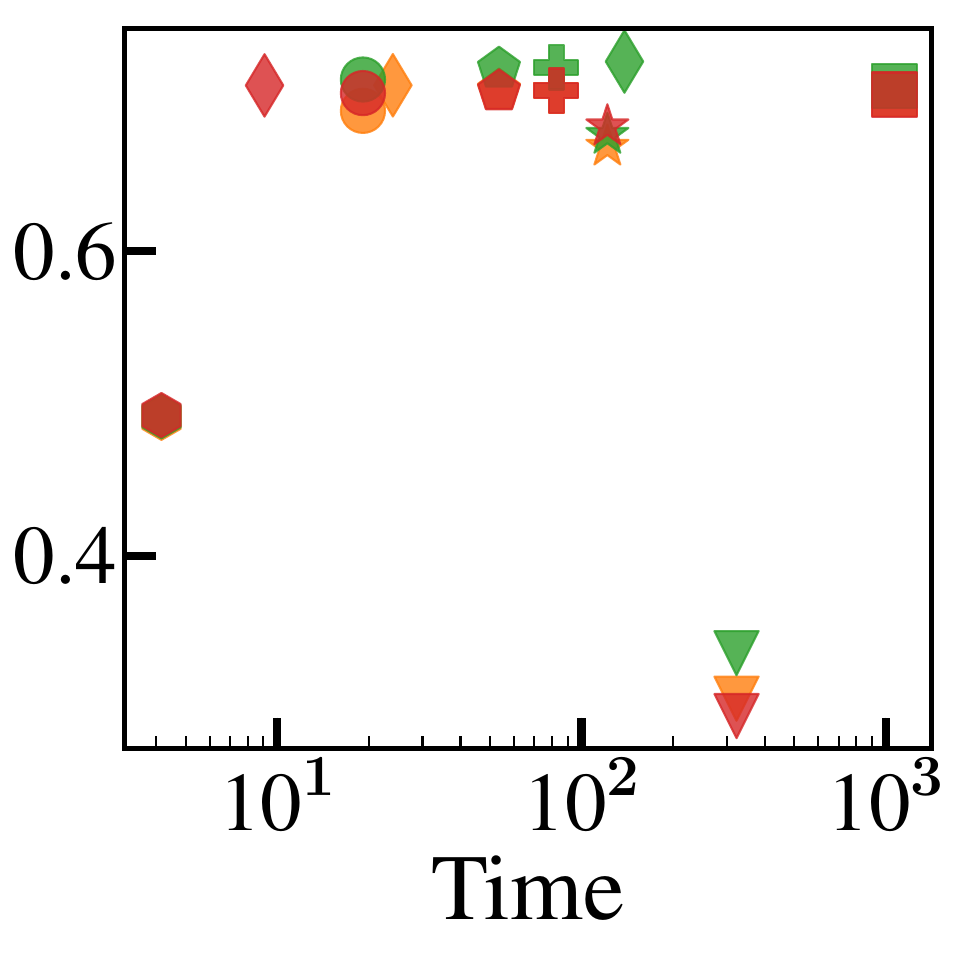}
					\caption{LPD}
				\end{subfigure}\hspace*{-0.5mm}%
				\hfill
				\begin{subfigure}[b]{0.23\textwidth}
					\centering
					\includegraphics[width=\linewidth]{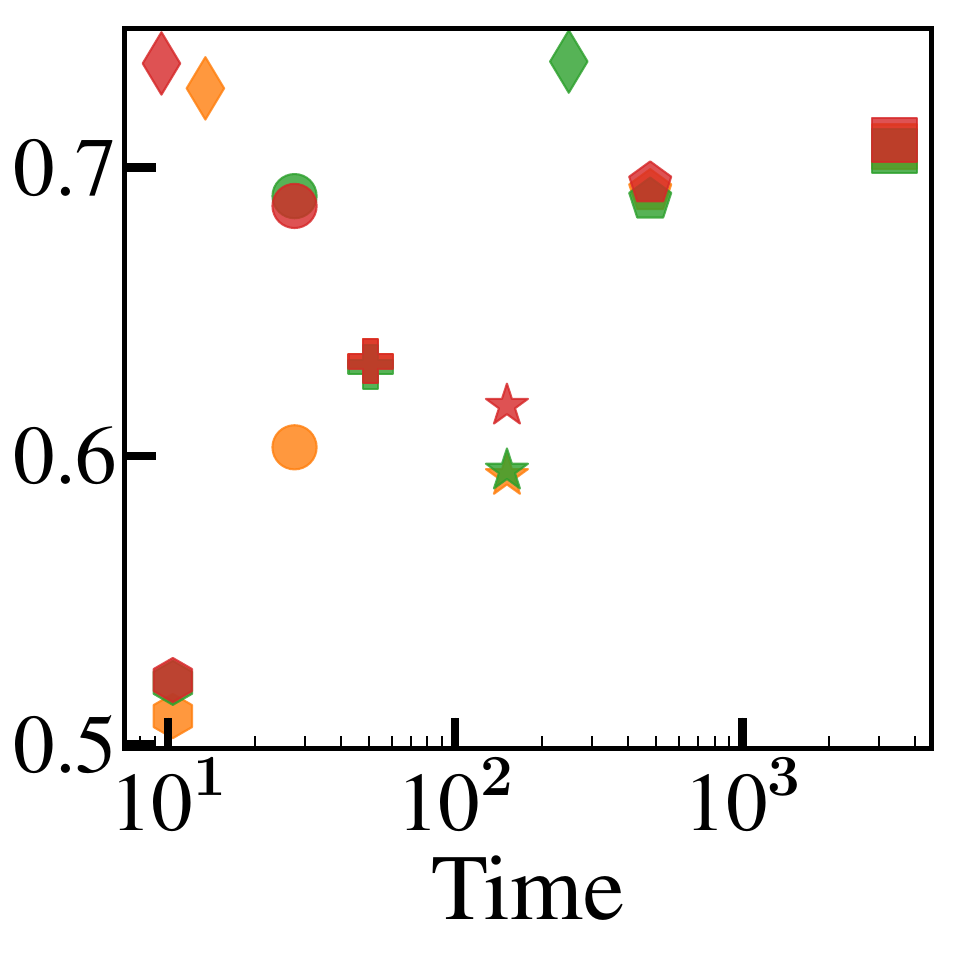}
					\caption{smoking}
				\end{subfigure}\hspace*{-0.5mm}%
			\end{minipage}
			
		\end{tabular}

        \caption{The comparison of accuracy–runtime trade-off. The different colors denote different downstream models, namely \textcolor[HTML]{FF7F0E}{\rule{0.8em}{0.4em}} (MLP), \textcolor[HTML]{D62728}{\rule{0.8em}{0.4em}} (SVM), and \textcolor[HTML]{2CA02C}{\rule{0.8em}{0.4em}} (FT-Transformer) and different shapes denote different baselines as shown in legend. }
        
        
		\label{fig:acc_time_overview}
	\end{figure*}

    \par In the second experiment, the efficiency of all methods is evaluated by runtime. Based on experiment results in Figure \ref{fig:time_overview_1x}, our DPDSyn achieves consistently better efficiency than comparable baselines. Specifically, on average, our DPDSyn achieves a maximum speedup of $160.40\times$, $333.73\times$, and $8.09\times$ in runtime on MLP, SVM, and FT-Transformer, respectively. The runtime of our DPDSyn is closely related to the used downstream model, and the running time of our DPDSyn is extremely small when the downstream task can be completed by simple AI models.
   
    \par To evaluate all methods more comprehensively, the trade-off between accuracy and runtime is illustrated in Figure \ref{fig:acc_time_overview}.  In particular, x-axis shows the mean of runtime of repeated experiments, and the y-axis shows the mean of accuracy. Ideally, overall performance of a method is better if it achieves high accuracy with low runtime. That is, when the representative point of a method is closer to the top-left corner, the overall performance of the method is better. According to the experiment in Figure \ref{fig:acc_time_overview}, our DPDSyn exhibits the optimal trade-off between accuracy and efficiency for MLP, SVM, and FT-Transformer models, thereby validating the excellent overall performance of DPDSyn. 

	\subsubsection{Scalability}


    \par To observe the scalability of all methods, the third experiment evaluates the accuracy of downstream models when the data volume increases. In particular, via bootstrap sampling, datasets' scale is expanded to 2× and 3× their original scale. Notably, the PrivMRF and AIM baselines are excluded from the evaluation due to experiment runtime exceeding 6 hours. The PrivPetal baseline is excluded because it fails when the dataset is scaled to 2× and 3× its original scale.
    

    \par According to experiment results in Table \ref{tab:Scalability}, DPDSyn exhibits superior scalability than other baselines across different scaling factors over all datasets. In particular, on average, on the 2× scaled datasets, DPDSyn achieves a maximum accuracy improvement of 1.53× in MLP, 1.50× in SVM, and 1.59× in FT-Transformer, compared to the benchmark baselines. On the 3× scaled datasets, DPDSyn achieves a maximum accuracy improvement of 1.53× in MLP, 1.44× in SVM, and 1.62× in FT-Transformer, compared to the benchmark baselines. The improvement of accuracy demonstrates the superior scalability of our DPDSyn. 

    \begin{table}[htb!]
		\centering
			\caption{ Accuracy Comparison across Scaled Datasets.}
			\label{tab:Scalability}
            \footnotesize
            \resizebox{\columnwidth}{!}{
			      \begin{tabular}{ccccccc}
				\toprule
				\multirow{2}{*}{Methods}     & \multicolumn{2}{c}{MLP}             & \multicolumn{2}{c}{SVM}             & \multicolumn{2}{c}{FT-Transformer}     \\ \cline{2-7}
				& \rule{0pt}{2.6ex}            2×               & 3×               & 2×               & 3×               & 2×               & 3×               \\  \midrule
                \textbf{NP Baseline}         & \textbf{0.7944}           & \textbf{0.8099}           & \textbf{0.7756}           & \textbf{0.7790}           & \textbf{0.8484}           & \textbf{0.8497}           \\ \midrule
				ABSyn                        & 0.7500           & 0.7489           & 0.7508           & 0.7499           & 0.7570           & 0.7507           \\
				PrivSyn                      & 0.5061           & 0.5107           & 0.5438           & 0.5446           & 0.4902           & 0.4836           \\
				MST                          & 0.7290           & 0.7299           & 0.7331           & 0.7359           & 0.7326           & 0.7310           \\
				MWEM                         & 0.7092           & 0.7201           & 0.7177           & 0.7222           & 0.7232           & 0.7275           \\
				DP\_GAN                      & 0.5243           & 0.5434           & 0.5167           & 0.5393           & 0.5167           & 0.5473            \\
				\textbf{DPDSyn}               & \textbf{0.7762}  & \textbf{0.7830}  & \textbf{0.7734}  & \textbf{0.7787}  & \textbf{0.7801}  & \textbf{0.7825}  \\ \midrule
				\textbf{Max. Imp.} & \textbf{1.53×} $\uparrow$ & \textbf{1.53×} $\uparrow$ & \textbf{1.50×} $\uparrow$ & \textbf{1.44×} $\uparrow$ & \textbf{1.59×} $\uparrow$ & \textbf{1.62×} $\uparrow$ \\ \midrule
				\textbf{Avg. Imp.} & \textbf{1.24×} $\uparrow$ & \textbf{1.24×}  $\uparrow$ & \textbf{1.22×} $\uparrow$ & \textbf{1.21×} $\uparrow$ & \textbf{1.26×} $\uparrow$ & \textbf{1.25×} $\uparrow$ \\ \bottomrule
			\end{tabular}
            }
	\end{table}

	\section{Conclusion}
    \par In this paper, we propose a novel method of DP dataset synthesis by introducing downstream task guidance. In particular, we train a DP AI model on the original private dataset such that the DP AI model preserves critical information for completing downstream tasks and the DP AI model is utilized to synthesize datasets such that the data utility is improved. Experimental evaluations demonstrate that our method achieves up to a 2.40× improvement in accuracy while reducing synthesis time by 333.73×. 

	\bibliographystyle{named}
	\bibliography{mybib/IJCAI_bibtex}
	
\end{document}